\documentstyle[11pt,newpasp,twoside,epsf,psfig,natbib]{article}
\def\edcomment#1{\iffalse\marginpar{\raggedright\sl#1\/}\else\relax\fi}
\reversemarginpar

\begin{document}
\title{Black hole binary mergers}
 \author{Junichiro Makino}
\affil{Department of Astronomy,
School of Science, The University of Tokyo,
7-3-1 Hongo, Bunkyo-ku, Tokyo 113-0033, Japan}

\hbadness 100000

\newcommand{\msun}{M_{\odot}}
\def\apgt{\ {\raise-.5ex\hbox{$\buildrel>\over\sim$}}\ }
\def\aplt{\ {\raise-.5ex\hbox{$\buildrel<\over\sim$}}\ }

\begin{abstract}

I overview the current understanding of the evolution
of massive black hole (MBH) binaries in the center of the
host stellar system.
One of the main questions is whether the stellar dynamical effect can
make the MBH binary hard enough that they can merge through
gravitational wave radiation. So far, theories and numerical
simulations suggested otherwize, since the hardening timescale becomes
very long once "loss cone" is depleted. I'll present the result of
recent simulations on this hardening timescale, and discuss its
implication on the formation history of massive black holes.

\end{abstract}

\section{Introduction}

There are now plenty of evidences
that many, if not most, of giant ellipticals contain massive central
black holes \citep{Magorrianetal1998}. Also, it has been suggested
that the black hole mass $M_{\rm BH}$ shows tight correlation with the
spheroidal mass and the central velocity dispersion
\citep{Gebhardtetal2000,FerrareseMerritt2000}. The most
straightforward explanation of such correlation is that massive
galaxies are formed by merging of less massive galaxies, and that the
central black hole also grows through merging
\citep{KauffmannHaehnelt2000}. This merging scenario naturally
explains the observed correlations.

This merging scenario has additional important characteristic that it
nicely explains the observed structure of the central region of
massive elliptical galaxies.  High-resolution observations by HST
revealed that the ``cores'' of the giant elliptical galaxies are not
really cores with flat volume density, but very shallow cusps
expressed as $\rho \propto r^{-\alpha}$, where $\rho$ is the volume
luminosity density and $r$ is the distance from the center, with power
index $\alpha = 0.5 \sim 1$ \citep{Gebhardtetal1996,Byunetal1996}.
\citet{MakinoEbisuzaki1996} performed the simulation of repeated
mergers of galaxies with central black holes. They found that the structure of the merger converges to
one profile. The merger product has a central cusp around the black
hole with slope approximately $-0.5$, and the total mass of the stars
in the cusp region is around the mass of the black hole
binary. \citet{NakanoMakino1999b,NakanoMakino1999a} showed, by
combination of simple $N$-body simulation and an analytic argument,
that this shallow cusp is explained by the fact that distribution
function of stars has a lower cutoff energy $E_0$. Their analytic
argument naturally explains the observed correlation between the cusp
radius and the black hole mass.

One remaining problem is what will happen to the binary black hole.
\citet[hereafter BBR]{Begelmanetal1980} predicted that the hardening
of the BH binary would be halted once it ejected all the stars in its
``loss cone''.  Early numerical simulations
\citep{Makinoetal1993,MikkolaValtonen1992} could not cover the range
of the number of particles (relaxation time) wide enough to see the
change in the timescale. Recent studies
\citep{Makino1997,QuinlanHernquist1997,Chatterjeeetal2003,MilosavljevicMerritt2001}
reported the results, which are not only  mutually inconsistent but
also inconsistent with the theoretical prediction. BBR predicted that
after the stars in the loss cone is depleted, the hardening timescale
would be the relaxation timescale of the parent galaxy, since in the
relaxation timescale stars would diffuse into the loss cone. The
results of numerical simulation ranges from no dependence on the
relaxation timescale \citep{MilosavljevicMerritt2001}, to some
dependence with upper limit in the timescale
\citep{QuinlanHernquist1997,Chatterjeeetal2003}, and finally to
weaker-than-linear dependence \citep{Makino1997}.

Thus, it has been an open question whether or not the loss-cone
depletion occurs. Some even argued that numerical simulations would
not help in determining the fate of a BH binary. Here, we present
the result of our resent $N$-body simulations with up to 1M particles,
where we saw a clear sign of the loss-cone depletion.

\section{Numerical method and  Initial Models}

The details of the model is given elsewhere\citep{MakinoFunato2003}.
We performed $N$-body simulations using direct summation method on
GRAPE-6\citep{Makinoetal2002}. Gravitational interaction between field
particles is softened, while that involves BH particles is not.

The initial galaxy model is a King model with nondimensional
central potential of $W_0=7$. We use the Heggie unit, where the mass
$M$ 
and the virial radius $R_v$ of the initial galaxy model and gravitational
constant $G$  are all unity.
The mass of black hole particles is
$M_{BH}=0.01$. They are initially placed at $(\pm 0.5,0,0)$ with velocity
$(0,\pm 0.1,0)$. 

The largest calculation (1 million particles and up to $t=400$) took
about one month, on a single-host, 4-processor-board GRAPE-6 system
with the peak speed of 4 Tflops.

\section{Results}

\begin{figure}
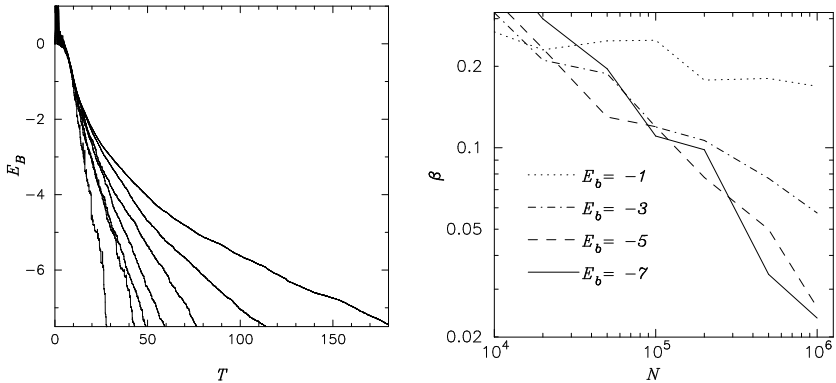

~ \psfig{file=ebfigmono.ps,width=5.1cm} ~~
 \psfig{file=defigmono.ps,width=5.4cm}

\caption{(Left) the evolution of the specific binding energy $E_b$ of the
binary black hole. Curves are result with $N=10K$, 20K, 50K, 100K,
200K, 500K and 1M (left to right). For all calculations, the softening
length is $\epsilon=0.01$. (Right)
the hardening rate $\beta = |\Delta E_b/\Delta t|$ plotted as a
function of the number of particles $N$. Dotted, dash-dotted, dashed
and solid curves denote the values measured from the time at which
$|E_b|$ reached 1, 3, 5 and 7, respectively. The end time is the time
at which $|E_b|$ reached initial value $+0.5$.
\label{fig:ebfign}}
\end{figure}

Figure \ref{fig:ebfign} shows the time evolution of the specific
binding energy (per reduced mass)  $E_b$ of the black hole binary. It is
clear that the hardening timescale becomes longer as we increase $N$
(and thereby the relaxation time).

To quantitatively evaluate the dependence of the hardening rate on the
number of particles, we calculated the hardening rate $\beta$ (see
figure 1).  When $E_b=-1$, the hardening rate is almost
independent of $N$. However, as the binary becomes harder, $\beta$
decreases, and the decrease is larger for larger $N$. Thus, the
hardening rate $\beta$ for large values of $|E_b|$ shows strong
dependence on the number of particles $N$. For $E_b=-7$, the hardening
rate $\beta$ is almost proportional to $1/N$. In other words, we
obtained the result which is consistent with the theory. The
hardening timescale is proportional to the relaxation timescale of the
parent galaxy.

\section{Discussions}

Our present result is consistent with the theory of loss-cone
depletion and its refilling in thermal timescale, while previous
numerical results are not. What caused this difference?

\citet{Makino1997} obtained the dependence much weaker than linear
from simulations with $N=2,048$ to 266,144. If we compare his figure 1
and our figure \ref{fig:ebfign}, the reason is obvious. The value of
the binding energy at which the dependence is measured was rather
small.  In other words, \citet{Makino1997} did not cover long enough
time.

\citet{QuinlanHernquist1997} performed simulations very similar to
the ones presented here. They found that the hardening rate was
practically the same for $N=10^5$ and $N=2\times 10^5$. We do not
really understand why they obtained this result. Thy used the SCF
method\citep{HernquistOstriker1992} and they varied the masses of
particles depending  on their initial angular momenta. We suspect this combination
might have complicated the dependence of the hardening timescale on
the number of particles. \citet{Chatterjeeetal2003} might have similar
problem.

\citet{MilosavljevicMerritt2001} performed three runs with 8K, 16K and
32K stars, and found that the hardening rate was independent of the
number of particles.  As suggested in their paper, this result is
simply because the loss cone was not depleted in their simulation,
primarily because $N$ was too small. So their result is not
inconsistent with ours.

As first suggested by BBR and confirmed by a number of followup works,
if the hardening timescale of the black hole binary is proportional to
the relaxation timescale of the parent galaxy, the evolution timescale
of a typical binary black hole in an elliptical galaxy exceeds the
Hubble time by many orders of magnitude. In other words, the binary is
unlikely to merge through  encounters with field
stars and gravitational wave radiation.

Our result strongly suggests that the hardening timescale is indeed
determined by the relaxation timescale, for large enough $N$ and after
the binary becomes sufficiently hard. Thus, our result imply that
gravitational interaction with field stars is insufficient to let the
binary merge.

If the binary black hole has long lifetime, it is quite natural
to assume that some of the host galaxies which contain binary black
holes would further merge with another galaxy with a central massive
black hole or a binary.
Simple estimate assuming the
thermal distribution of the eccentricity \citep{MakinoEbisuzaki1994}
suggests that, during repeated three body interactions, the
eccentricity of the binary can reach a very high value, resulting in 
quick merging through gravitational wave radiation. Thus, strong
triple interaction of three supermassive black holes may be common.

\def\mn{MNRAS}
\def\nat{Nature}


\newcommand{\noopsort}[1]{} \newcommand{\printfirst}[2]{#1}
  \newcommand{\singleletter}[1]{#1} \newcommand{\switchargs}[2]{#2#1}

\end{document}